\documentclass[12pt]{iopart}
\usepackage{iopams}  
\usepackage{graphicx}
\usepackage{cite}
\usepackage{url}

\begin{document}

\title[]{Emergence of a finite-size-scaling function in the supervised learning of the Ising phase transition}

\author{Dongkyu Kim and Dong-Hee Kim}

\address{Department of Physics and Photon Science, Gwangju Institute of Science and Technology, Gwangju 61005, Republic of Korea}
\ead{dongheekim@gist.ac.kr}

\begin{abstract}
We investigate the connection between the supervised learning of 
the binary phase classification in the ferromagnetic Ising model
and the standard finite-size-scaling theory of 
the second-order phase transition.
Proposing a minimal one-free-parameter neural network model, 
we analytically formulate the supervised learning problem
for the canonical ensemble being used as a training data set.
We show that just one free parameter is capable enough to describe 
the data-driven emergence of the universal finite-size-scaling function 
in the network output that is observed in a large neural network, 
theoretically validating its critical point prediction 
for unseen test data from different underlying 
lattices yet in the same universality class of the Ising criticality.
We also numerically demonstrate the interpretation with the proposed
one-parameter model by providing an example of finding a critical point 
with the learning of the Landau mean-field free energy being applied to
the real data set from the uncorrelated random scale-free graph
with a large degree exponent.
\end{abstract}

\maketitle

\section{Introduction}
\label{sec:intro}

Understanding how an artificial neural network learns the state of matter 
is an intriguing subject for the applications of machine learning 
to various domains including the study of phase transitions in 
physical systems~\cite{Carleo2019,Ohtsuki2020,Zdeborova2020,Carrasquilla2020,Bedolla-Montiel2020}. 
In a typical form of the multilayer perceptron, a neural network consists of 
layers of neurons that are connected through a feedforward network structure. 
The network can produce a mathematical function approximating 
any desired outputs for given inputs in principle
\cite{Cybenko1989,Hornik1989,Hornik1991,Leshno1993}, 
and one can optimize associated neural network parameters 
for a particular purpose in a data-driven way. 
In the supervised learning for the classification of data, 
which we particularly focus on in this study,
the network is optimized to reproduce the labels of the already classified 
training input data. 
Remarkably, the neural network trained in a data-driven way often produces
prediction with some accuracy even for unacquainted data
of a similar type, which is not necessarily 
from the same data set or system given in the training.
With various machine learning schemes being examined, 
the phase classification and the detection of a phase transition point 
have been extensively studied in classical and quantum model systems 
in recent years~\cite{Carrasquilla2017,Nieuwenburg2017,Wang2016,Ohzeki2016,Ohtsuki2017,Tanaka2017,Hu2017,Wetzel2017a,Wetzel2017b,Ponte2017,Suchsland2018,Kim2018,Iso2018,Huembeli2018,Liu2018,Beach2018,Vargas-Hernandez2018,Mills2018,Morningstar2018,Li2018,Kashiwa2019,Zhang2019a,Casert2019,Zhang2019b,Greitemann2019a,Liu2019,Greitemann2019b,Liu2020,Rao2020,Kiwata2019,Efthymiou2019,Li2019,Dong2019,Canabarro2019,Gannetti2019,Lee2019,Shiina2020,Blucher2020,DAngelo2020,Munoz-Bauza2020,Veiga2020}.

Because it is data-driven, rather than being based on the first principles, 
witnessing the empirical successes naturally 
leads to fundamental questions such as what specific information
the neural network learns from the training data, 
to what extent and why it works even for unacquainted data or systems,
how trustworthy such data-driven prediction can be,
and most importantly, what is the mathematical foundation of 
the learnability. 
A general difficulty in addressing these questions is due to the nature of 
the ``black box'' model where one can hardly see inside because
of high complexity generated by the interplay between a large number of 
neural network components.
While the opaque nature may not harm its empirical usefulness especially
when it works as a recommender, transparency can be crucial in 
the applications requiring extreme reliability where one wants 
logical justification of how it reaches such predictions.
Explainable machine learning to deal with issues along this direction 
has attracted much attention in domains of scientific 
applications~\cite{Roscher2020}. 
In the machine-learning detection of phase transitions and critical phenomena, 
there are increasing efforts in interpreting how the machine prediction 
works or designing transparent machines such as demonstrated
in several previous studies 
\cite{Ponte2017,Kim2018,Suchsland2018,Casert2019,Zhang2019b,Greitemann2019a,Liu2019,Greitemann2019b,Liu2020,Rao2020,Blucher2020}.
Our goal in this paper is to interpret the predicting power of
a neural network classifying the phases of the Ising model 
into the conventional physics language of the critical phenomena 
by proposing an analytically solvable model network having just one free
parameter.

The Ising model has been employed as a popular test bed of 
machine learning and is particularly useful for our purpose
of discussing the learnability
since it is a well-established model of the second-order phase transition
in statistical physics.
Our work is closely related to the seminal work by 
Carrasquilla and Melko~\cite{Carrasquilla2017} where a network with 
a single hidden layer of $100$ neurons was trained with 
the pre-assigned phase labels of the Ising spin data
that were given accordingly whether the data is sampled below or above
the known critical point of the training system.
It turned out that the one trained for the square lattices 
was reusable to the unseen data from the triangular lattices 
without any cost of a new training, providing a good estimate of 
a critical point with a finite-size-scaling behavior. 
In our previous work~\cite{Kim2018}, we investigated 
this reusability by downsizing the neural network. 
We found that the hidden layer could be as small as the one 
with just two neurons without loss of the prediction accuracy. 
In the downsized network model, 
we argued that its reusability to the systems in the different lattices 
is encoded in the system-size scaling behavior of 
the network parameters which is universal for any other lattices
in the same universality class.

In this paper, we further simplify the neural network model 
into a minimal setting with just one free parameter, providing
a more transparent mathematical view on how the learning and 
prediction of the critical point occurs with the data of the Ising model.
Despite the minimal design, we find that a single parameter is only
necessary to capture the behavior observed in a large neural network 
that plays an essential role in the predicting accuracy 
and the reusability acquired from the training.
The present one-parameter model improves the idea of 
the previous neural network models~\cite{Carrasquilla2017,Kim2018}
in terms of transparency and analytical interpretability.
In our previous two-node model~\cite{Kim2018} that needs
two free parameters, the fluctuations of the order parameter were ignored 
for the convenience of analytic treatment, which we find is important and 
now fully incorporated into the present derivation with the one-parameter model.
On the other hand, the three-node model proposed previously 
by Carrasquilla and Melko~\cite{Carrasquilla2017} 
also had one free parameter but unfortunately was not analytically 
explored further. While the third hidden neuron is unnecessary
in our model, our derivations and discussions can be 
directly applied to the previous three-node model because of 
the similarity between their functional forms of the output.

Analytically minimizing the cross entropy for 
the supervised learning with the canonical ensemble
at an arbitrarily large system size, 
we show that the trained network output becomes
a universal scaling function of the order parameter
with the standard critical exponents.
This emergence of the scaling function is consistent with
the empirical observation in a large neural network,
which we find works as a universal kernel for the prediction with 
unseen test data from different lattices but belonging to 
the same Ising universality class.
We demonstrate the operation of the one-parameter model
by presenting the learning with the Landau mean-field free energy
and its prediction accuracy of the critical point with the data 
from the uncorrelated random scale-free graph that belongs to 
the mean-field class.

This paper is organized as follows. In Sec.~\ref{sec:method},
the procedures of the supervised learning are described. 
In Sec.~\ref{sec:univfunc}, the implications of the scaling form 
emerging in the network output are discussed. 
In Sec.~\ref{sec:main}, the one-parameter neural network is presented
with the derivation of the analytic scaling solution.
The demonstration with the Landau mean-field 
free energy and the application to the data of the Ising model on 
the random scale-free graph is given in Sec.~\ref{sec:example}. 
The summary and conclusions are given in Sec.~\ref{sec:conclusions}. 

\section{Supervised learning of the phase transition in the Ising model}
\label{sec:method}

We consider the classical ferromagnetic spin-$1/2$ Ising model 
with the nearest-neighbor exchange interactions 
without a magnetic field, which is described by the Hamiltonian
$\mathcal{H} = -J \sum_{\langle i,j \rangle} s_i s_j$
where the spin $s_i$ at a site $i$ takes the value of 
either $1$ or $-1$, and the summation runs over all the nearest-neighbor 
sites $\langle i,j \rangle$ in the given lattices. 
The interaction strength $J$ and the Boltzmann constant $k_\mathrm{B}$ 
are set to be unity throughout this paper. 
The spin configuration $\mathbf{s}\equiv\{s_1,s_2,\ldots,s_N\}$ 
is given as an input to train the neural network, which is labeled as 
the ordered or disordered phase depending on whether the temperature 
associated with the data is lower or higher than the critical 
temperature $T_c$ given for the supervision.
The learning with the labeled data for the binary 
classification can be done by minimizing the cross entropy~\cite{book1,book2}, 
\begin{equation}
    \mathcal{L}(\mathbf{x}) = - \sum_{\mathbf{s}} \left[ 
    Q(\mathbf{s})\ln F(\mathbf{s};\mathbf{x}) 
    + \left(1-Q(\mathbf{s})\right)
    \ln \left(1 - F(\mathbf{s};\mathbf{x})\right) 
    \right] \,,
\end{equation}
with respect to the neural network parameters $\mathbf{x}$. 
The function $Q(\mathbf{s})$ returns the binary value 
$0$ or $1$ representing the label of the data $\mathbf{s}$.
The function $F(\mathbf{s};\mathbf{x})$ is the output of 
the neural network, giving a value between $0$ and $1$
for an input $\mathbf{s}$. The parameter $\mathbf{x}$ 
is to be optimized to maximize the likelihood between 
the distribution of the output $F$ and the given distribution 
of the actual label $Q$. 

We prepare the data set of spin configurations at a given 
temperature by assuming an unbiased sampling with the Boltzmann 
probability in the canonical ensemble.
The unbiased sampling is important to the mechanism of
predicting a correct $T_c$ with the trained network.
While our main results are obtained from the analytic calculation 
of the cross entropy minimization based on our one-parameter model of
the neural network, we also need the numerically 
generated data for the verification with real lattice geometries.
Depending on the necessities in the numerical demonstration, 
we employ the Wang-Landau sampling method~\cite{WL1,WL2,Landau2004}
computing the joint density of states or the Wolff cluster 
update~\cite{Wolff1989} generating the spin configuration data.

The prediction of a critical point is done based on 
how the network output behaves with the temperature
associated with the test data given as an input. 
However, it comes with practical ambiguity arising from the fact
that the value of the output fluctuates severely across 
the inputs at the temperatures near the critical point. 
While one might consider a smooth curve of an average 
$\langle F \rangle$ evaluated over many test inputs 
at a given temperature,
one would still need a criterion or threshold 
to discriminate the order and disorder phases.
There are previously suggested ways to obtain the location of 
a transition point, such as the scheme of the learning 
by confusion~\cite{Nieuwenburg2017}. 
In the simplest case, where the training data thoroughly 
covers very fine grids of temperature across the transition
as we consider here,
one can just pick a certain cut such as $\langle F \rangle = 1/2$ 
used in Ref.~\cite{Carrasquilla2017} to get the estimate of 
a transition temperature.
Interestingly, the temperature corresponding to the cut 
showed a finite-size-scaling behavior with various sizes of 
the systems being examined~\cite{Carrasquilla2017},
and it turned out that a specific value of the cut would not
matter in the finite-size-scaling analysis~\cite{Kim2018}.

\section{Emergence of the universal scaling function in the network output}
\label{sec:univfunc}

\begin{figure}
    \centering
    \includegraphics[width=\textwidth]{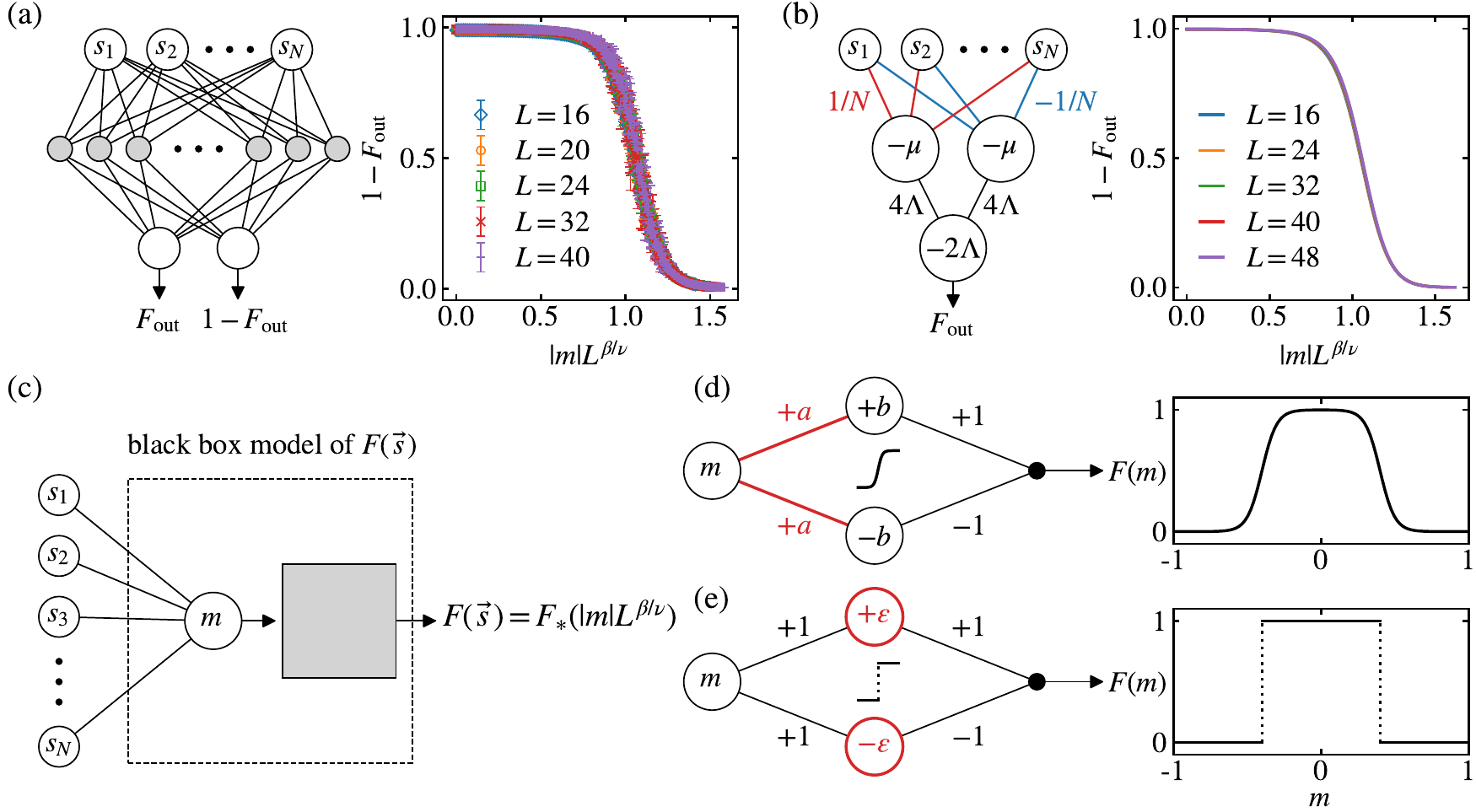}
    \caption{Neural network models for the phase classification
    in the Ising model. The output $F$ of the trained network 
    is plotted as a function of the order parameter $m$ computed 
    for every individual  input data of the spin configuration.
    (a) The network model with a single hidden layer of many neurons 
    \cite{Carrasquilla2017}, which is trained here with $50$ hidden neurons
    (see Ref.~\cite{Kim2018} for the detail of the data preparation). 
    The marker and error bar indicate the average and range of 
    the output at each $m$, respectively.
    (b) The previous two-node model trained with 
    the Wang-Landau data~\cite{Kim2018}. 
    (c) A schematic diagram of the structure producing 
    the universal scaling function. 
    Minimal models of the gray box in (c) are sketched in (d) and (e)
    with the sigmoid and Heaviside step activation functions, respectively.}
    \label{fig1}
\end{figure}

Revisiting the behavior of the previous large-size neural network 
trained with many hidden neurons~\cite{Carrasquilla2017}, we observe
a particular scaling form emerging in the network output when it is 
plotted as a function of the order parameter $m = \sum_i s_i / N$ 
for every individual data $\mathbf{s}$.
Figure~\ref{fig1}(a) shows the training results with the data 
in the square lattices of $N = L \times L$ sites, and
it turns out that these input-output curves for various system sizes 
fall onto a common curve in the scaling of $|m|L^{\beta/\nu}$ with 
the critical exponents $\beta$ and $\nu$ of the Ising universality class
in two dimensions.
Our previous two-node model~\cite{Kim2018} shows the same feature 
in the network output as shown in Fig.~\ref{fig1}(b).

The emergence of the scaling function $F_*(|m|L^{\beta/\nu})$ 
in the network output reveals the simple explanation of 
how it finds a genuine critical point and why it works even
for unseen test data generated in different lattices yet in 
the same Ising universality class of the critical exponents.
For instance, two different trainings on the square and triangular 
lattices would give the neural networks with exactly the same scaling 
form of the output if it is plotted as a function of $m$.

With the test inputs of $\mathbf{s}$ being sampled from the probability 
distribution $p_L(\mathbf{s},T)$ at a temperature $T$ 
in the system of size $L$, the averaged network output is written as
$\langle F_L \rangle = \int p_L(\mathbf{s},T) F_L(\mathbf{s}) \,\mathrm{d}\mathbf{s}$.
If the test data set is prepared unbiasedly from the canonical ensemble, 
then the test data distribution can be expressed 
by the finite-size-scaling form~\cite{Binder1981,Bruce1981,Nicolaides1988} of 
$p_L(\mathbf{s},T) \equiv L^{\beta/\nu} p_*(|m|L^{\beta/\nu},t L^{1/\nu})$
near the critical point $T_c$, 
where $t \equiv T/T_c - 1$ denotes the reduced temperature.
Consequently, going across the critical point in the temperature axis, 
the averaged network output is finally rewritten as
\begin{equation} \label{eq:gfunc}
    \langle F_L \rangle = \int \mathrm{d}m L^{\beta/\nu} 
    p_*(|m|L^{\beta/\nu},t L^{1/\nu}) 
    F_*(|m|L^{\beta/\nu}) \equiv G_*(t L^{1/\nu}) \,,
\end{equation}
which is exactly what was observed numerically 
in the previous work~\cite{Carrasquilla2017}. 

The function $G_*(tL^{1/\nu})$ immediately indicates that 
the crossing point at $t=0$ between the curves of different $L$'s 
gives an exact critical point $T_c$ that is associated with $p_*$ of 
the \emph{test} data set.
In Ref.~\cite{Carrasquilla2017},
the temperatures corresponding to $\langle F_L \rangle = 1/2$ 
were extrapolated toward an infinite $L$ to predict $T_c$. 
Equation~(\ref{eq:gfunc}) explicitly shows that it is 
equivalent to following a constant $G_*(tL^{1/\nu})$ that leads 
to the line of $T_L = T_c + a L^{-1/\nu}$, where
the specific value of $1/2$ does not play any role. 
Therefore, the key feature that guarantees the physically meaningful
$T_c$ prediction is whether or not the neural network 
properly approximates the scaling form $F_*$ of the output function. 
While the numerical training can be affected in practice 
by the detail of the data preparation such as 
the temperature grid spacing, the accurateness of the $F_*$ form 
determines the quality of the training in this particular learning problem 
with the data of the Ising model. 

We emphasize that $F_*$ and $p_*$, the two constituents of
Eq.~(\ref{eq:gfunc}), correspond to the two different data sets 
of the \emph{training} and \emph{testing} systems, respectively.
Because the validity of Eq.~(\ref{eq:gfunc}) requires the same critical 
exponents for both of the training and testing data sets,
this condition precisely defines the limit of the applicability,
indicating that the prediction works for unseen data only
in a particular group of the systems of the same criticality.
For instance, as previously demonstrated in Ref.~\cite{Kim2018}, 
the neural network trained for the Ising model in the square lattices 
fails to predict a correct transition point 
for the data of the three-dimensional lattices. 
If the test and training systems are not in the same universality
class of the Ising critical exponents, a crossing point between 
the $\langle F_L \rangle$ curves cannot be properly identified, 
and the extrapolation of $T_L$ at a fixed $\langle F_L \rangle$ 
gives a different $T_c$ depending on a choice of 
the value of $\langle F_L \rangle$. 

An important question is then how the neural network becomes 
approximating the universal kernel $F_*(|m|L^{\beta/\nu})$ 
in the supervised learning of the binary phase labels. 
The functional form suggests that the network is trained to read 
the order parameter from the input, which is consistent with 
the previous observations~\cite{Carrasquilla2017,Wetzel2017b,Kim2018}.
Thus, we may be able to consider a picture that is schematically 
shown in Fig.~\ref{fig1}(c), 
where $m$ is assumed to be transmitted with the trivial link weights 
$1/N$ from the input to the hidden layer that belongs to the gray box. 
In the following section, we present a minimal neural network 
model of the gray box to show how the critical behavior of 
the training data leads to such scaling form of 
the output function.

\section{One-free-parameter neural network model}
\label{sec:main}

\subsection{Previous two-node model and further simplification}

In our previous work~\cite{Kim2018}, we introduced a model network 
with the downsized hidden layer of just two hidden neurons  
receiving the explicit order parameter and demonstrated that 
it did not lose any predicting power and accuracy.
In Fig.~\ref{fig1}(b), we verify that it indeed produces 
the expected form $F_*$ of its output, explaining the high accuracy 
of the critical point predicted by this downsized network
in the previous work.
However, despite the simple structure that allows analytic 
treatment to some extent, our previous two-node 
model is not mathematically transparent enough to see  
how the exact form of $F_*(|m|L^{\beta/\nu})$ emerges
from the learning.

The technical difficulties in our previous analytic approach stem 
from the use of the sigmoid activation function assigned to 
both of the hidden and output neurons.
While this is a common setting for a usual large-size
neural network to be trained for a binary classifier, 
it leads to an output function that can be written as 
\begin{equation}
    F_L(m) = f[ 4\Lambda f(m - \mu_L) + 4\Lambda f(-m-\mu_L) - 2\Lambda_L ],
\end{equation}
where $f(x) = \frac{1}{2}(1+\tanh\frac{x}{2})$ is the sigmoid function,
which is plotted in Fig.~\ref{fig1}(b) with the parameters trained
in the square lattices. 
We previously derived the system-size scaling behavior of 
the two neural parameters as 
$\mu_L \sim L^{-2\beta/\nu}$ and $\Lambda_L \sim L^{2\beta/\nu}$ 
by ignoring the order parameter fluctuation in the input data set,
which was a crude assumption since the fluctuations of $m$
are severe near the critical point.
With careful approximations with expansions for $m$ around $F_L=1/2$, 
the form of $F_*(|m|L^{\beta/\nu})$ might be justifiable, but
a simpler and intuitive model would be certainly preferred
to provide a more transparent picture of the learning process.

Thus, we present a minimally simple one-parameter model of the gray box 
in Fig.~\ref{fig1}(c) generating a very simple step-wise output function, 
\begin{equation} \label{eq:Fout_ideal}
    F(m;\epsilon) = \Theta(m+\epsilon) - \Theta(m-\epsilon),
\end{equation}
where $\Theta(x)$ is the Heaviside step function.
The corresponding neural network structure is sketched in 
Fig.~\ref{fig1}(e).
The one with the sigmoid neuron given in Fig.~\ref{fig1}(d) is 
its differentiable version, which is equivalent to the Heaviside one 
in the limit of large $a$ and $b$ with $\epsilon \equiv b/a$ being finite. 
Note that the output neuron simply reduces the signal from the hidden layer
without any activation function and bias being involved.
It is trivial to see that the wanted form of $F_*(|m|L^{\beta/\nu})$
would appear if the free parameter $\epsilon$ is given as 
$\epsilon_L \propto L^{-\beta/\nu}$, 
which we show below indeed occurs in the supervised learning 
with the cross entropy minimization.

\subsection{Scaling solution of the free parameter}

The binary phase label of the data is expressed as 
$Q(T) \equiv \Theta(T-T_c)$ with the critical point $T_c$ being 
given for the supervision.
The training data set is represented by 
the probability distribution function $p_L(m,T)$ of the order 
parameter $m$ at a temperature $T$ in the system of size $L$.
The temperature range of the data set  
can be given as $T \in [T_l, T_h]$ where $T_l \ll T_c \ll T_h$. 
The cross entropy is then rewritten as
\begin{equation} \label{eq:cross_entropy}
    \fl
    \mathcal{L}(\epsilon) = 
    - \int_{T_l}^{T_h} dT \int_{-\infty}^\infty dm \, 
    p_L(m,T) \left[ Q(T) \ln F(m,\epsilon)
    + [1-Q(T)] \ln [1 - F(m,\epsilon)] \right] .
\end{equation}

For the mathematical convenience, we first employ a differentiable 
version of the output function $F$ shown in Fig.~\ref{fig1}(d) that is written as
\begin{equation}
   F(m, \epsilon \equiv b/a) =  \frac{1}{2} \left[ \tanh \frac{a(m+\epsilon)}{2}
   - \tanh \frac{a(m-\epsilon)}{2} \right].
\end{equation}
While the two parameters $a$ and $b$ appear in this expression,
it is effectively a one-parameter model because the output function 
essentially depends on the ratio $\epsilon \equiv b/a$ in the limit
of large $a$ and $b$ that we assume.
Taking the derivative of $\mathcal{L}$ with respect to $\epsilon$
in the limit of large $a$, we obtain an integral equation,
\begin{equation} \label{eq:Lmin}
    \int_{T_c}^{T_h} \mathrm{d}T \int_\epsilon^{\infty} 
    \mathrm{d}m \, p_L(m,T)
    = \int_{T_l}^{T_c} \mathrm{d}T \int_0^\epsilon 
    \mathrm{d}m \, p_L(m,T) ,
\end{equation}
where we assume that $p_L$ is an even function of $m$
as we consider the unbiased preparation of the training data set 
preserving the Ising symmetry.
Equation~(\ref{eq:Lmin}) is solvable for the system-size scaling
behavior of $\epsilon$ under ideal training conditions 
where the training data is in the canonical ensemble and 
uniformly available at all temperatures. 
While such training data set is typically considered 
in the Monte Carlo simulations, it also allows a fully analytic treatment
based on the standard finite-size-scaling ansatz of $p_L$
near the critical point $T_c$.

While $T_h$ and $T_l$ can be given to be arbitrarily far from $T_c$, 
the specific values of $T_h$ and $T_l$ are unimportant 
because $\int_\epsilon^\infty p_L(m,T)\,\mathrm{d}m$ 
is only meaningful in the critical area.
In the right hand side of Eq.~(\ref{eq:Lmin}), 
$p_L$ is sharply peaked at $|m|=1$ deep in the ordered phase ($T \ll T_c$),
leading to a negligibly small value of $\int_0^\epsilon p_L \,\mathrm{d}m$
if $\epsilon$ is much less than one. On the other hand,
in the left hand side, at $T \gg T_c$ in the disordered phase,
$p_L$ is governed by the central limit theorem, and then the integral 
$\int_\epsilon^\infty p_L \,\mathrm{d}m$ decays
asymptotically as $\exp(-N\epsilon^2)/(\sqrt{N}\epsilon)$ 
if $\sqrt{N}\epsilon$ increases with the number of spins $N$.
Therefore, presumed that $\epsilon$ decreases with 
the system size $L$ while $\sqrt{N}\epsilon$ increases with $L$,
we can replace $T_l$ and $T_h$ with effective bounds
of the critical area. The width of the critical area decreases 
with increasing $L$, suggesting that one must consider 
the finer grids of temperature for the data 
of the larger system in a numerical approach.
In the analytic calculations,  all temperatures 
are available in the training data set.

Considering the temperature integration
over the critical area of $t \in [-\delta t, \delta t]$, 
where $t \equiv (T-T_c)/T_c$ denotes the reduced temperature,
the probability distribution function $p_L(m, T)$ near $T_c$ 
can be expressed as 
$p_L \equiv L^{\beta/\nu}p_*(mL^{\beta/\nu},t L^{1/\nu})$
by the standard finite-size-scaling ansatz. 
Then, Eq.~(\ref{eq:Lmin}) can be rewritten as 
\begin{equation} \label{eq:fss}
    \int_0^{\delta t L^{1/\nu}} \mathrm{d}\tau 
    \int_{\epsilon L^{\beta/\nu}}^\infty \mathrm{d}x \,
    p_*(x,\tau) = 
    \int_{-\delta t L^{1/\nu}}^0 \mathrm{d}\tau 
    \int^{\epsilon L^{\beta/\nu}}_0 \mathrm{d}x \,
    p_*(x,\tau) ,
\end{equation}
where the change of variables is performed for 
$x \equiv mL^{\beta/\nu}$ and $\tau \equiv tL^{1/\nu}$.
This equation holds for an arbitrarily large system
to be trained. With the scale invariance of the equation 
for an arbitrary $L$ being imposed, the width of the critical area
is clarified to be $\delta t \sim L^{-1/\nu}$, 
and more importantly, the neural parameter
has to behave as $\epsilon \sim L^{-\beta/\nu}$. 
This system-size scaling behavior of $\epsilon$ directly leads to 
the expected form of the network output $F_*(|m|L^{\beta/\nu})$ 
that we have discussed above. 

One can verify that the resulting scaling solution of 
$\epsilon \sim L^{-\beta/\nu}$ indeed validates the assumption that 
$\sqrt{N}\epsilon$ would increase as $L$ increases,
which we have used in the derivation.
In $d$ dimensions, it is rewritten 
as $\sqrt{N}\epsilon \sim L^{d/2-\beta/\nu}$.
It is easy to see that the condition $(d/2 - \beta/\nu) > 0$ 
holds from the hyperscaling relations that indicates
its equivalence to the critical exponent $\gamma > 0$ 
of the susceptibility divergence.

Instead of using the differentiable version of $F$ and taking 
the limit of a infinite $a$, one can also introduce a small 
shift $0 < \delta \ll 1$ directly to Eq.~(\ref{eq:Fout_ideal}) 
to avoid the undefined evaluations of $\ln F$ and $\ln (1-F)$ as
\begin{equation}
    F(m,\epsilon) = (1-\delta)[\Theta(m+\epsilon) - \Theta(m-\epsilon)] 
    + \delta \,,
\end{equation}
which can be trivially implemented in the model sketched in Fig.~\ref{fig1}(e)
by adjusting the link weights and the bias of the output neuron.
After $\ln\delta$ is factored out, the cross entropy can be
rewritten as
\begin{equation}
    \frac{\mathcal{L}(\epsilon)}{2|\ln\delta|} = 
    \int_{T_c}^{T_h} \mathrm{d}T 
    \int_\epsilon^\infty \mathrm{d}m \, p_L(m,T)
    +\int_{T_l}^{T_c} \mathrm{d}T 
    \int_0^\epsilon \mathrm{d}m \, p_L(m,T) \, ,
\end{equation}
where $\delta$ does not affect the optimization. 
Following the procedures that we have shown above, 
the minimization of the cross entropy can then be rewritten 
with the temperature integration range effectively being limited to 
the area around a given $T_c$ as
\begin{equation}
    \int_0^{\tau_o} p_*(\epsilon L^{\beta/\nu},\tau) \, \mathrm{d}\tau =  
    \int_{-\tau_o}^0 p_*(\epsilon L^{\beta/\nu},\tau) \, \mathrm{d}\tau \, , 
\end{equation}
where $\tau_o \equiv \delta t L^{1/\nu}$ denotes the effective
width of the critical area normalized with the system size.
The scale-invariant solution of this equation that holds 
for an arbitrarily large $L$ provides the same finite-size-scaling
behavior of $\epsilon \sim L^{-\beta/\nu}$ and thereby produces 
the universal kernel of the $F_*$ function of the network output.

\begin{figure}
    \centering
    \includegraphics[width=0.65\textwidth]{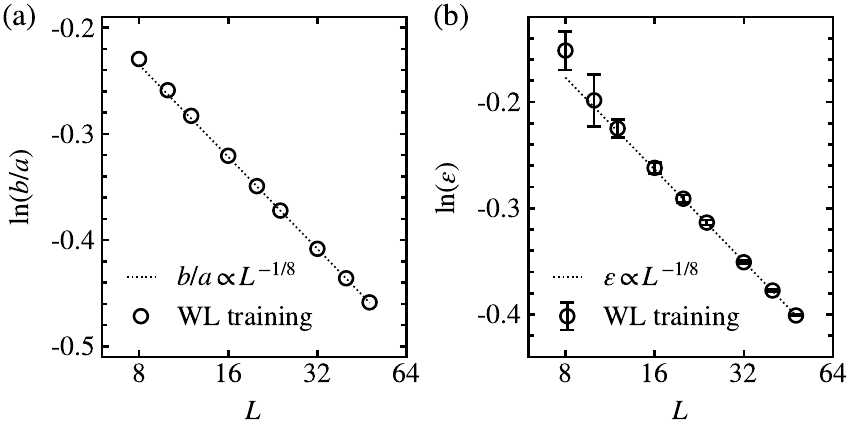}
    \caption{Parameters of the minimal neural network models 
    determined by training with the Wang-Landau data in the square lattices.
    The scaling behavior is plotted as a function of linear 
    dimension $L$ of the training system for (a) the ratio $b/a$ 
    of the model in Fig.~\ref{fig1}(d) and 
    (b) the parameter $\epsilon$ of the model in Fig.~\ref{fig1}(e).}
    \label{fig2}
\end{figure}

In addition, we numerically verify the derived scaling solution 
of $\epsilon \sim L^{-\beta/\nu}$ in the system on square lattices. 
The input data for the training is provided from the estimate
of the probability distribution $p_L(m,T)$ that is directly given
by the Wang-Landau sampling of the joint density of states
\cite{WL1,WL2,Kim2018}. Since the Wang-Landau estimate of $p_L$ 
provides unlimited access to all temperatures, one can numerically 
evaluate Eq.~(\ref{eq:cross_entropy}) and perform the minimization.
Figure~\ref{fig2} shows the system-size scaling of the trained 
parameter $b/a$ for the model of Fig.~\ref{fig1}(d)
and $\epsilon$ for the model of Figs.~\ref{fig1}(e), respectively.
Because $m$ is discrete in a finite system, 
the use of the Heaviside step function causes a range of $\epsilon$ 
corresponding to the same value of the cross entropy.
The error bars in Fig.~\ref{fig2}(b) present such ranges of $\epsilon$,
showing that the degeneracy diminishes as the system size gets larger.
Both of the two numerical training results show excellent agreement 
with our derivation of $\epsilon \sim L^{-\beta/\nu}$ 
with the critical exponent $\beta/\nu = 1/8$ being in the universality 
class of the two-dimensional Ising model. 

\subsection{Fluctuations in the input layer extracting the order parameter}

\begin{figure}
    \centering
    \includegraphics[width=0.65\textwidth]{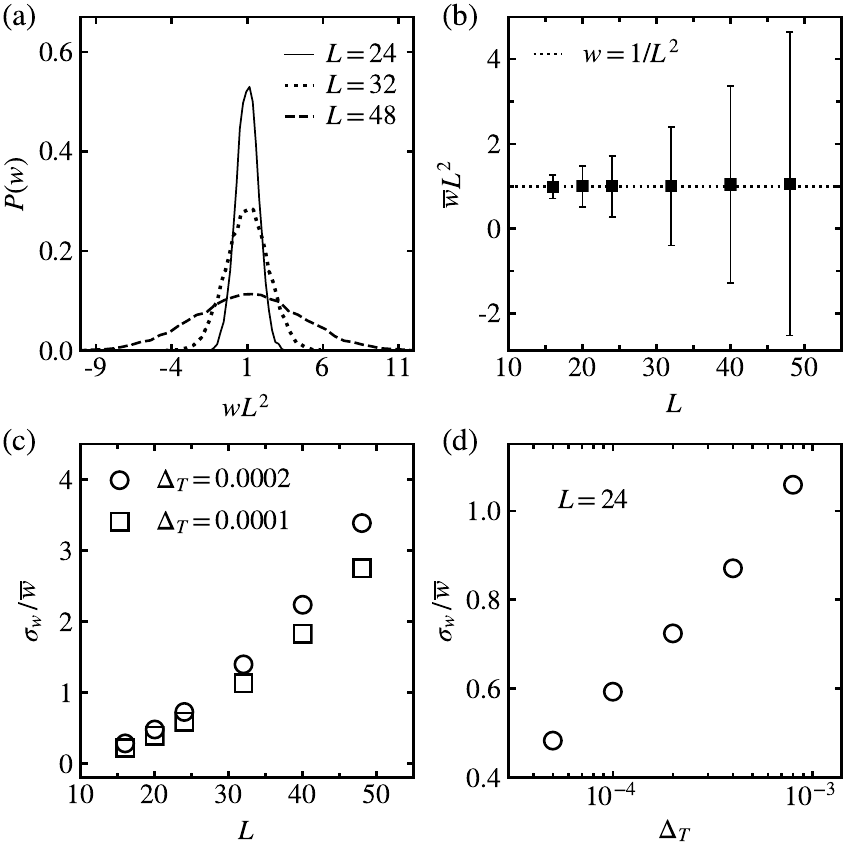}
    \caption{Weight fluctuations of the input links extracting 
    the order parameter. The statistics of the link weights $w$ connecting
    input and hidden layers is examined by using the stochastic training 
    in the model of the Heaviside neurons in Fig.~\ref{fig1}(e)
    with the data of spin configurations in the square lattices.
    (a) The system-size dependence of the link weight distribution 
    $P(w)$ for a given temperature spacing at $\Delta_T = 0.0001$ 
    of the Monte Carlo training data.
    (b) The average $\overline{w}$ (symbols) and standard deviation 
    $\sigma_w$ (error bars) plotted as a function of the system size $L$ 
    for the training data prepared with $\Delta_T = 0.0001$.
    The panels (c) and (d) indicate that the fluctuations of $w$ decrease
    as the temperature spacing $\Delta_T$ gets smaller.}
    \label{fig3}
\end{figure}

The minimal network model presented above has the designed 
input layer with the fixed link weight $w_i = 1/N$,
transmitting the explicit order parameter $m=\sum_i w_i s_i$ 
from the input of the spin configuration $\mathbf{s} \equiv \{s_i\}$. 
This design provides the simplest model that corroborates 
the observation from the large-scale network trained 
under null hypothesis~\cite{Carrasquilla2017,Kim2018}. 
In a practical point of view, discarding the uncertainty 
of the link weights helps to remove the training noises 
and the numerical overfitting as implied 
in the comparison of the output functions 
between Fig.~\ref{fig1}(a) and Fig.~\ref{fig1}(b), 
reducing the error in the $T_c$ estimate eventually.
Still, it is an interesting question to ask how stable 
this ideal model of the input weight $w_i = 1/N$ would be 
if variations on $w_i$ are allowed in the training and 
also how the fluctuations of $w_i$ would depend 
on the detail of the training data preparation.

Introducing an additional set of undetermined parameters 
$\mathbf{w} = \{w_i\}$ for $m = \sum_i w_i s_i$
in the minimal model, we obtain $w_i$ by solving the equation 
$\left[\frac{\partial \mathcal{L}}{\partial \mathbf{w}}\right]_{\epsilon = \epsilon_L} = 0$
with the parameter $\epsilon$ being fixed at the ideal training 
solution that we have already obtained from 
$\left[\frac{\partial \mathcal{L}}{\partial \epsilon}\right]_{(w_i=1/N)} = 0$.
We employ the stochastic gradient method in the scheme of the online 
learning~\cite{Bottou1998} in combination with the Wolff cluster update 
algorithm to sample the data of spin configurations in the square lattices.
The temperature grids of the training data are set in the range of 
$[T_c/2, 3T_c/2]$ with the spacing of $\Delta_T$.

Figure~\ref{fig3} displays how the probability distribution of 
the weights depends on the system size and the temperature grid spacing 
of the training data prepared. 
It turns out that the resulting distribution $P(w)$
is bell-shaped with a well-defined average at the ideal value 
of $\bar{w} = 1/N$. The ratio of the standard deviation and the average 
of $P(w)$ represents the magnitude of fluctuations in $w_i$, 
which increases as the system size gets larger
but decreases as the temperature grid space $\Delta_T$ 
gets smaller.

This observation hints a proper data preparation for 
a more accurate prediction of $T_c$ in the numerical training. 
The observed behavior of $P(w)$ implies that the training
of the larger system would need the finer grids in 
temperature for the training data set to suppress the noises 
in the final form of the network output $F*(|m|L^{\beta/\nu})$ 
as it affects the accuracy as a $T_c$ locator.
This test indicates the importance of the thorough coverage of 
the critical area in the training data set, 
which in some sense makes the machine learning less magical but 
is reasonable in terms of statistical physics 
because finding a genuine critical point cannot be separated from 
the critical behavior of the system.

\section{Learning the Landau mean-field theory of the Ising model}
\label{sec:example}

For the demonstration of how the prediction works on unseen data 
from a different underlying geometry, 
we choose the Landau mean-field free energy as a generator
of the training data and then apply it to the test data 
produced on a scale-free graph as an underlying geometry 
of the Ising model. 
By using the analytically trained one-parameter network model, 
we attempt to locate the critical point of the Ising model 
in the random scale-free graph with a large degree exponent 
that is known to be in the mean-field class~\cite{Dorogovtsev2002,Goltsev2003,Hong2007}.

For the order parameter $m$ with the Ising symmetry, 
the Landau mean-field free energy per spin can be written 
at $T$ near the critical point $T_c$ as 
\begin{equation}
\label{eq:LandauFE}
    f(m,t) = f_0 + a_2 t m^2 + a_4 m^4 \, ,
\end{equation}
where $t \equiv (T-T_c)/T_c$ denotes the reduced temperature, 
and $a_2$ are $a_4$ are positive constants.
In the system of $N$ spins, the corresponding probability distribution of 
the order parameter $m$ near $T_c$ is written directly 
from the Landau free energy as
\begin{equation}
    p_N(m,t) \propto N^{1/4} \exp [ - N(a_2 t m^2 + a_4 m^4)/T_c ] \, , 
\end{equation}
which leads to the finite-size-scaling form,
\begin{equation}
   p_N(m,t) = N^{1/4} p_*(mN^{1/4}, tN^{1/2}) \, .
\end{equation}
One can verify the mean-field exponents $\beta = 1/2$ and $\bar{\nu} = 2$ 
in the comparison with the standard finite-size-scaling ansatz
$p_N = N^{\beta/\bar{\nu}} p_*(mN^{\beta/\bar{\nu}},tN^{1/\bar{\nu}})$.
We do not consider any possibility of the logarithmic corrections, 
and we strictly limit our demonstration in the class 
of systems that is described  by the Landau free energy in Eq.~(\ref{eq:LandauFE}). 

Provided the scaling form of the probability distribution $p_N(m,t)$ 
of a training data set, it is now straightforward to obtain 
the scaling behavior of the parameter $\epsilon$ in the trained network.
Following the analytic minimization of the cross entropy 
given in Sec.~\ref{sec:main}, we can write down 
the scaling solution as $\epsilon_N = \epsilon_0 N^{1/4}$
by just putting $N^{\beta/\bar{\nu}}$ instead of $L^{\beta/\nu}$. 
The constant $\epsilon_0$ can be determined by the detail of $p_N(m,t)$ 
and the range of temperature of the training data set.
However, a particular value of $\epsilon_0$, which we just set to be 
the unity in the calculation below, 
is unimportant to the performance of the neural network 
because it does not affect the form of the output function 
$F_N(m) = F_*(|m|N^{\beta/\bar{\nu}})$ that works as a universal kernel.

\begin{figure}
    \centering
    \includegraphics[width=0.65\textwidth]{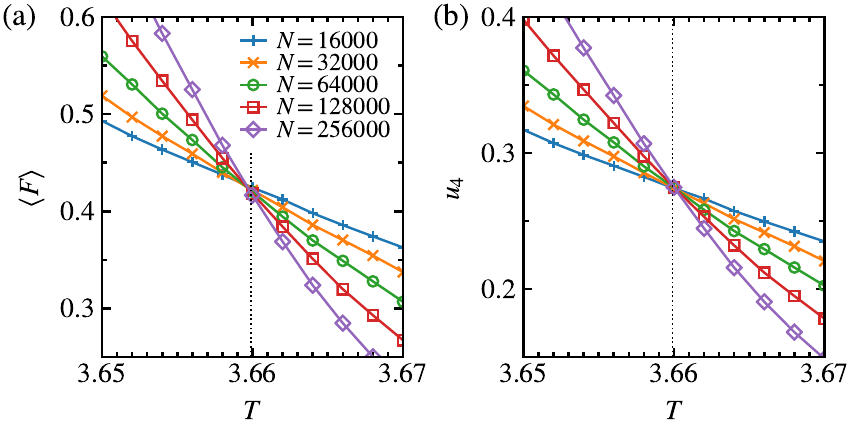}
    \caption{Critical point detection by applying the mean-field-trained neural 
    network to the data of the Ising model on the uncorrelated scale-free graph 
    with the degree exponent $\gamma=6.5$. (a) The network output $\langle F \rangle$
    averaged over the inputs of Monte Carlo data sampled at each temperature.
    (b) The fourth-order cumulant as a function of temperature given for comparison.
    Each data point is an average over the ensemble of 50,000 random graph 
    samples, and the error bars (not shown) are smaller than the marker size.
    The vertical dotted lines indicate the exact location of the critical point.}
    \label{fig4}
\end{figure}

We examine the accuracy of the $T_c$ prediction with the data from
the uncorrelated  random scale-free graph model~\cite{Catanzaro2005}. 
In the graph of underlying vertices of the Ising spins, 
a vertex is randomly connected to some number of other vertices 
by the edges representing the exchange interactions between the residing spins.
The number of the edges from a vertex, referred to as the degree,
follows a power-law distribution $p(k) \sim k^{-\gamma}$ with a degree 
exponent $\gamma$. It is known that when $\gamma > 5$, the Ising model
on this scale-free graph exhibits the mean-field critical exponents 
\cite{Dorogovtsev2002,Goltsev2003,Hong2007}.
Here we examine the scale-free graph with the degree exponent 
$\gamma = 6.5$ and the minimum degree 4.
The degree of each site is given as the greatest integer less 
than or equal to the value drawn randomly from the power-law distribution.
The test input data set consists of 10,000 spin configurations 
per a graph sample obtained from the Wolff cluster updates 
at each temperature, and 50,000 random graph samples are included
in the test data set.

Figure~\ref{fig4} shows the output of the one-parameter model 
averaged over the test inputs of the Ising spin 
data prepared on the scale-free graph geometry. 
The crossing point between the curves of different system sizes
provides the estimate of the critical point $T_c = 3.6595(5)$, 
which is in good agreement with the standard detection 
using the fourth-order cumulant that gives $T_c = 3.6601(2)$. 
The exact critical point for the uncorrelated random scale-free 
graph with $\gamma > 5$ was derived previously in Ref.~\cite{Dorogovtsev2002} as
$T_c = 2/\ln[ \langle k^2 \rangle /(\langle k^2 \rangle - 2 \langle k \rangle)]$,
which becomes $T_c \simeq 3.6599$ for our degree sequences 
of the scale-free graph samples examined.

These results demonstrate the operation of Eq.~(\ref{eq:gfunc}) 
in the mean-field regime with the training of the Landau free energy 
and the Monte Carlo test data generated on the random scale-free graph.
In the previous work~\cite{Kim2018}, we did similar tests
with the two-node model that has two free parameters, presenting
the cases where the training and test data sets are in
the same and different universality classes. The validity
of the $T_c$ prediction is only guaranteed when the training
and test data sets are in the same universality class
while the underlying geometries are not necessarily the same.
We argue that Eq.~(\ref{eq:gfunc}) is the simple physical basis 
that explains the valid $T_c$ prediction with the supervised learning 
on the Ising model, which can be implemented 
by using just a single free parameter of the neural network.

\section{Conclusions}
\label{sec:conclusions}

We have investigated the connection between 
the data-driven prediction of a critical point based on 
the supervised learning in the ferromagnetic Ising model
and the standard finite-size-scaling theory of the second-order 
phase transition. 
It turns out that the scaling form $F_*(|m|L^{\beta/\nu})$ emerging 
in the network output is the source of the predicting power,
which works as a universal kernel guaranteeing a physically legitimate 
estimate of a critical point for unseen test data from different lattices 
but in the same universality class with the training data. 
We have shown that a minimal network with just one free parameter
suffices to model such emergence of the scaling form 
in the minimization of the cross entropy.
For the training data unbiasedly sampled
from the canonical ensemble, we have derived the analytic 
scaling solution of the one-parameter model that leads to 
the expected scaling form of the output function.
For the numerical demonstration, we have considered the Landau
mean-field energy as a generator of the training data
and verified that it accurately locates the critical point
on the random uncorrelated scale-free graph  
that belongs to the mean-field class.

While we have shown that the conventional finite-size-scaling ansatz 
can be implemented in a very simple data-driven way 
with just one parameter being learned, the model benefits
from the simple order parameter structure of the Ising model
that is extracted easily as empirically observed 
in the large-size neural network.
The possible direction for future studies may include
the generalization for more complex symmetry of an order 
parameter and the interpretation of the learning in a broader
range of phase transitions and critical phenomena 
in complex systems.

\ack
This work was supported from the Basic Science Research Program through the National Research Foundation of Korea funded by the Ministry of Science and ICT (NRF-2019R1F1A1063211) and also from a GIST Research Institute (GRI) grant funded by the GIST.

\end{document}